\documentclass[11pt]{article}
\usepackage[dvips]{epsfig} 
\usepackage{times} 
\usepackage{picinpar}
\usepackage{wrapfig}
\usepackage{floatflt}
\makeatletter
\renewcommand{\section}{\@startsection{section}{1}{0in}
        {0.4\baselineskip}{0.1\baselineskip}{\Large\bf}}
\renewcommand{\subsection}{\@startsection{subsection}{2}{0in}
        {0.25\baselineskip}{-\baselineskip}{\large\bf}}
\renewcommand{\subsubsection}{\@startsection{subsubsection}{3}{0in}
        {0.1\baselineskip}{-\baselineskip}{\normalsize\bf}}
\makeatother
\begin{document}
\vspace{2ex}
%
\thispagestyle{myheadings}
%
\markright{OG.2.1.08}
\begin{center}
%
{\LARGE \bf VHE Spectral Properties of Mrk~501 with the CAT telescope}
\end{center}

\begin{center}
%
%
{\bf J.P. Tavernet$^{1}$ for the C{\small AT} Collaboration}\\
{\it $^{1}$L.P.N.H.E. Paris 6/7 - 4,Place Jussieu - 75252 Paris Cedex 05 -
  France}
\end{center}

\begin{center}
{\large \bf Abstract}
\end{center}
We report here observations of the active galactic nucleus Mrk~501, at
energies above 250 GeV carried out with the CAT atmospheric imaging telescope
from March 1997 to Autumn 1998. This source was in a high state of activity at
several different wavelengths in 1997, and the observed flux at TeV energies
has been seen to change by a factor of $\sim$ 20 from from 1995 and 1996 fluxes. CAT
observations also indicate a curved spectrum at TeV energies, and a correlation
between the gamma-ray intensity and the spectral hardness. The temporal
variability and the TeV spectral properties are examined.
\vspace{-0.5ex}
%
\vspace{1ex}

\section{Introduction:}
\label{intro.sec}
The active galactic nucleus (AGN) Mrk~501 is one of the closest (z=0.034) BL
Lacert{\ae}  objects.  It was discovered at TeV energies by the Whipple
Observatory (Quinn et al., 1996), and confirmed by HEGRA (Bradbury et al.,
1997) and the C{\small AT} collaboration (Punch et al., 1997).

From March to October 1997, VHE observations of Mrk~501 revealed extreme
variability, with a measured flux up to 8 times the Crab Nebula flux. Several
independent \v Cerenkov telescopes have confirmed this dramatic activity
(Protheroe et al., 1998). Based on observations carried out in 1997, we present
a detailed study of spectral properties. Thanks to the large variability of the 
source during this period, the spectral hardness was studied as a function of 
the intensity of the source and spectra were derived for different intensity levels. 

Mrk~501 has been regularly observed by the C{\small AT}  telescope
   since March 1997 and the resulting light curve is presented.

\section{C{\small AT} \v Cerenkov Imaging Telescope and data analysis}
\label{cat.sec}
Located on the Th\'emis solar plant in southern France (2$^\circ$E, 42$^\circ$N,
1650 m a.s.l.), the C{\small AT} (\v Cerenkov Array at Th\'emis) imaging
telescope has been described in detail in Barrau et al. (1998). It achieves a 
low threshold ($\sim$ 250 GeV) despite its small
reflector area (17.7 m$^2$) by taking full advantage of the rapidity of the \v
Cerenkov pulse with a near-isochronous mirror, fast phototubes, fast trigger and
readout electronics. Moreover, its very high definition camera (546 pixels
with an angular size of 0.12$^\circ$) allows an accurate analysis of the
longitudinal and lateral light profile of the shower image, as discussed in Le
Bohec et al. (1998), giving a good separation of gamma-ray showers from hadronic 
ones by means of a $\chi^2$-like variable and of the pointing angle $\alpha$.
   
Cuts used in the standard C{\small AT} analysis are the following:
 $\hbox{P}\left(\chi^2\right) > 0.35$, $\alpha$ $\le$ 6$^\circ$,
Q$_\mathrm{tot}$ (the total charge in the image) $\ge$ 30 $\gamma$e ; 
an additional cut requiring at least 3 $\gamma$e in the fourth-brightest 
pixel ensures uniform trigger conditions. These cuts yield a background 
rejection factor of about 200, with a $\gamma$-ray efficiency better than 40\%.

Observations are done in an ON/OFF mode: the source is tracked (ON) for $\sim$
30 minutes and, to estimate the background, the telescope tracks a 
control region offset in right ascension on the source. 
Typically, one OFF-region is observed for every two ON-regions.
Data are further selected for good sky-conditions and stable detector
operation.

\section{Source Variability and Light Curve}
\label{variab.sec}

Here, we include data from Spring 1997 to Autumn 1998, representing a total of $\sim$
100 hours of observation. The data-set has been further restricted to zenith
angles $< 45^\circ$, for which the detector calibration studies have been
completed.  Figure 1 shows the Mrk~501 nightly integral flux levels
above 250 GeV.

\begin{figwindow}[1,r,%
{\mbox{\epsfig{file=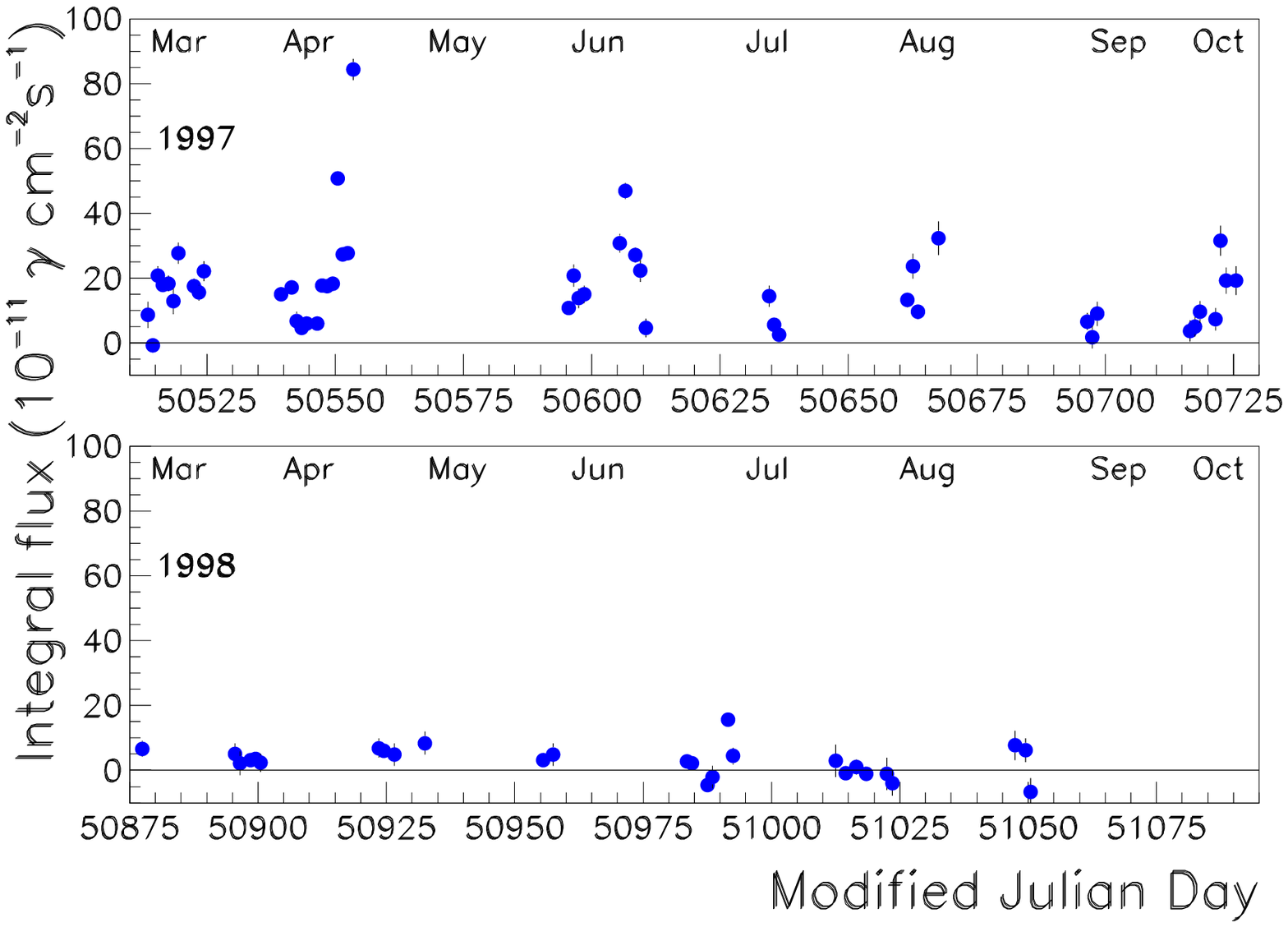,width=4.0in}}},%
{Mrk~501 nightly integral flux levels above $250\:\mathrm{GeV}$,
expressed in units of  
$10^{-11}\:\mathrm{cm^{-2}s^{-1}}$, for observations in 1997. The
average spectral shape as derived in Sect.~4 has been assumed.\\}]
In 1997, VHE $\gamma$-ray emission was very
strong and significant variability was observed in the nightly averages.  The
highest flare, recorded on April 16$^\mathrm{th}$, 1997 reached about 8 times the
Crab level. On the other hand in 1998, the integral flux appears to have been fairly
constant and low ($\sim$ $\frac{1}{10}$ of Crab nebula rate).

While variability at a nightly scale is directly seen in Figure 1, the search
for intra-night variability has been studied in detail in Renault C., Renaud
N. and Henri G. (1998). No significant short-term variability was found.\\
\end{figwindow}

\vspace{2.3cm}

\section{Differential Energy Spectra}
\label{spect.sec}

The effective detection area and the energy resolution 
function are derived from simulations as functions of energy and zenith 
angle. These simulations take account of the detector response and are 
calibrated on the basis of the large statistics of gamma-ray excess 
events from runs on Mrk~501 with a high signal-to-noise ratio, and of 
\v Cerenkov rings induced by muons.
\begin{figwindow}[1,r,%
{\mbox{\epsfig{file=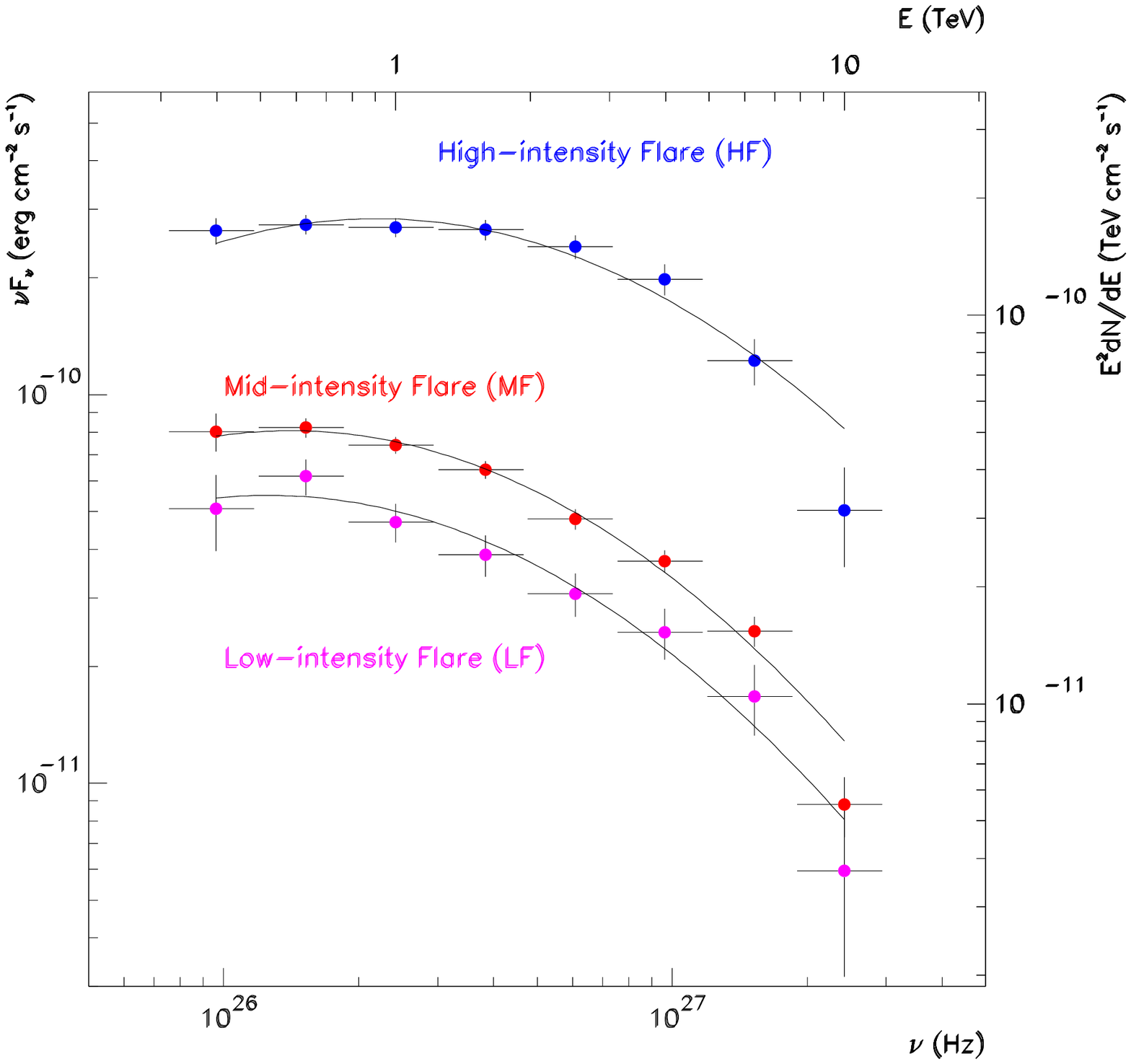,width=4.35in}}},%
{$\nu F_\nu$ or $E^2\,\mathrm{d}\Phi/\mathrm{d}E$ spectra for the HF, MF, and LF states.\\
Individual intensities per bin are only indicative.\\}]
The differential energy spectra are obtained, for a given hypothesis on the
spectral shape, by a maximum-likelihood fit taking into account the effective
area and the energy-resolution function of the telescope.
Two possible spectral shapes have been investigated: a pure power law
($\mathcal{H}_0$) and a curved shape ($\mathcal{H}_1$) defined by $\phi_0
E_\mathrm{TeV}^{-(\alpha + \beta\log_{10}\!E_\mathrm{TeV})}$, suggested by
general considerations on $\gamma$-ray emission from blazars. The likelihood
ratio of the two hypotheses, defined as $ \lambda = -2 \times \log
\frac{\mathcal{L}(\mathcal{H}_0)}{\mathcal{L}(\mathcal{H}_1)} $, gives an
estimate of the relevance of $\mathcal{H}_0$ with respect to $\mathcal{H}_1$.\\

For this spectral analysis, the data-set was taken from March to October 1997.
During this period, three different activity states of Mrk~501 have been
defined: {\it HF} for high-intensity flares with an integral flux 
($\ge 50 \times 10^{-11}$ cm$^{-2}$s$^{-1}$), {\it LF} for low-intensity runs with an 
integral flux ($\le 12 \times 10^{-11}$ cm$^{-2}$s$^{-1}$) 
and \\ {\it MF} for mid-intensity flares 
with an integral flux between {\it HF} and {\it LF}.
\noindent
The average over the complete data-set (AV) has also been considered.
Results for each data-subset are given in Table~\ref{tabspresults}.
Only statistical errors are quoted in this table.

The resulting differential energy spectra of Mrk~501 are shown for energies 
from 330 GeV up
to 10 TeV in Figure 2. The likelihood ratio, $\lambda$, 
behaves like a $\chi^2$ with one degree of freedom. Thus, from the values of $\lambda$, 
the probabilities of falsely accepting
$\mathcal{H}_1$, over $\mathcal{H}_0$, are $7\times 10^{-8}$ and $7\times
10^{-12}$ respectively, if the power-law hypothesis were true. 
The weak intensity of the source in the LF state may
explain the lower significance of the curvature term $\beta$ in this state. 
The VHE peak emission of Mrk~501 clearly takes
place in the range of several hundred GeV. The corresponding peak 
energies, $E_\mathrm{max}$, in the spectral energy distribution 
($E^2 \mathrm{d}\phi/\mathrm{dE}$) have been derived from the fitted values of $\alpha$ 
and $\beta$ (Table 1).
\end{figwindow}
There is some indication of an increase in
$E_\mathrm{max}$ with source intensity, which is equivalently shown by the
variation of the fitted values of $\alpha$, corresponding to a possible
correlation between intensity and $E_\mathrm{max}$.  The intensity-hardness
correlations, discussed in the next section, support this effect.
Moreover, a correlated variability in X-ray and $\gamma$-ray have been 
observed (Djannati-Atai et al., 1999). So, together with the power deficit in the GeV range
(Samuelson et al., 1998), 
this strongly suggests a two-component emission spectrum for Mrk~501 (Fossati et al., 1998).
\begin{table*}[!ht]
\begin{center}
\caption{The differential flux is given in units of $10^{-11}\:\mathrm{cm^{-2}s^{-1}TeV^{-1}}$. 
In the hypothesis of a curved shape ($\mathcal{H}_1$), it is parametrized as 
$\mathrm{d}\phi/\mathrm{d}E_\mathrm{TeV}\,= 
\phi_0 E_\mathrm{TeV}^{-(\alpha + \beta\log_{10}\!E_\mathrm{TeV})}$. $\phi_0^\mathrm{pl}$ 
and $\alpha^\mathrm{pl}$ refer to a pure power-law hypothesis ($\mathcal{H}_0$).
The fit has been performed in the interval from 
$330\:\mathrm{GeV}$-$13\:\mathrm{TeV}$. 
The last columns give the peak-emission energies $E_\mathrm{max}$.}
\label{tabspresults}
\begin{tabular}{llccccccc}
\hline
\noalign{\smallskip}
Set & ON  & $\phi_0^\mathrm{pl}$ &$\alpha^\mathrm{pl}$ & $\phi_0$
&$\alpha$ & $\beta $& $\lambda $& $E_\mathrm{max}$ \\ 
 & (h)   &  &  & &  &  &  & (GeV) \\ 
\noalign{\smallskip}
\hline
\hline
\noalign{\smallskip}
LF&  13.6 & $\;\:2.72 \pm 0.13$ & $2.45 \pm 0.05 $ 
  & $\;\:3.13 \pm 0.19$ & $2.32 \pm 0.09$& $0.41 \pm 0.17$&
10.7 & $410 \pm 201 $ \\
MF&  40.5 & $\;\:4.10 \pm 0.10$ & $2.46 \pm 0.03$  
  & $\;\:4.72 \pm 0.14$ & $2.25 \pm 0.05$ & $0.52 \pm 0.08$ & 
   47.1& $583\pm104$ \\
HF& $\;\:3.1$ & $14.5\;\: \pm 0.50$ & $2.21 \pm 0.03$ 
  & $17.6\;\: \pm 0.61$ & $2.07 \pm 0.04$ & $0.45 \pm 0.09$ &
  29.1 & $840\pm 108$  \\
\hline
AV&  57.2 & $\;\:4.56 \pm 0.10$ & $ 2.46\pm 0.02$ 
  & $\;\:5.19 \pm 0.13$ & 
  $2.24 \pm 0.04$ & $0.50 \pm 0.07$ & 61.5&  $578\pm\;\; 98$  \\
\noalign{\smallskip}
\hline
\end{tabular}
\end{center}
\end{table*}

\section{Source Intensity vs. Spectral Hardness}
\label{hard.sec}

The hardness ratio is defined as $R = \frac{N_{E>E_\mathrm{mid}}}
{N_{E>E_\mathrm{low}}}$, where $N_{E>E_\mathrm{low}}$ (resp. $N_{E>E_\mathrm{mid}}$) 
is the number of events with a fitted energy greater than $E_\mathrm{low}$ 
(resp. $E_\mathrm{mid}$).

All data have been divided into five sets with different average
fluxes. For each data-set the hardness ratio has been computed for three 
different energy bands,
$[E_\mathrm{low},E_\mathrm{mid}]_\mathrm{GeV}
=\{[450,900];[600,1200];[900,1500]\}$,
hereafter referred to as $R_{[>900/>450]}$, $R_{[>1200/>600]}$, and
$R_{[>1500/>900]}$. 
This method is more robust than the usual spectrum analysis for the following 
reasons:\\
 - since data have been divided into two bins in energy, statistical fluctuations 
   on the single parameter {\it R} are reduced;\\
 - $E_\mathrm{low}$ is always chosen well-above the detector threshold in order to
    have a good estimation of the energy;\\
 - the data set is limited to zenith angles $<25^\circ$ in order to avoid large 
   threshold variations;
   
\begin{figwindow}[1,r,%
{\mbox{\epsfig{file=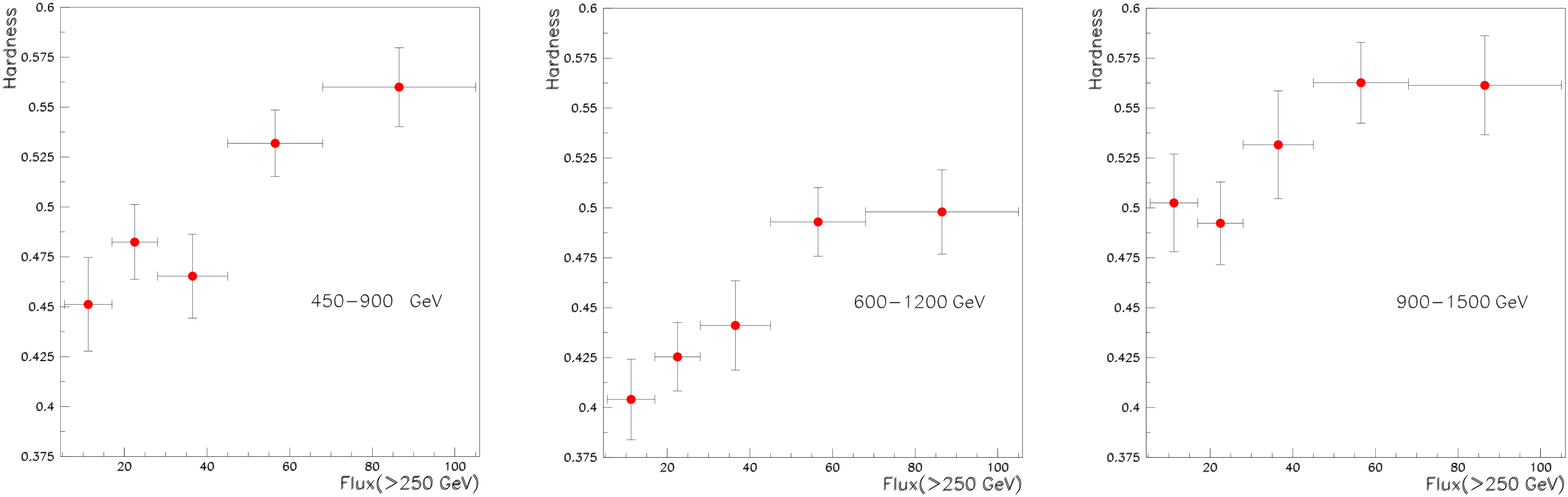,width=16.7cm}}},%
{ Hardness-ratio {\it vs.} source intensity for three different energy bands. 
Intensities are given as the integral flux above $250\:\mathrm{GeV}$ in units 
of $10^{-11}\:\mathrm{cm^{-2}s^{-1}}$.\\}]
\noindent
 - finally, an additional cut on the shower impact parameter (required 
   to be $< 130$m) allows the energy resolution to be improved.\\
\end{figwindow}

 The hardness ratio for the three different energy bands is shown in
 Figure 3. A correlation between hardness and intensity is clearly
 observed for all energy-bands. The $\chi^2$ probabilities that the
 distributions are flat for $R_{[>900/>450]}$, $R_{[>1200/>600]}$ and
 $R_{[>1500/>900]}$ are: $3.8\times 10^{-4}$, $9.0\times
 10^{-4}$, and $7\times 10^{-2}$, respectively.\\[0.3cm]

\section{Conclusions}
\label{conc.sec}

The emission from Mrk~501 in 1997 and 1998 has been monitored with the CAT
telescope.
The spectrum has been measured from 330 GeV up to 10 TeV in
1997, revealing several spectral properties. A curved spectral shape 
has been derived, confirming the curvature reported by the Whipple group (Samuelson et al., 1998)
and HEGRA collaboration (Aharonian et al., 1999).
The emission is seen to extend above 10 TeV. The observed intensity-hardness 
correlation can be simply ascribed to a shift of the peak TeV emission energy. 
This is confirmed by the observed increase in $E_\mathrm{max}$ derived from the spectral 
analysis.

\vspace{1ex}
\begin{center}
{\Large\bf References}
\end{center}
%
Aharonian et al. 1999, to be published in A\&A.\\
Barrau, A., et al. 1998, Nucl. Instr. Meth. A416, 278\\
Bradbury, S., et al., 1997, A\&A 320, L5 \\
Djannati-Atai, A. et al., 1999, submitted to A\&A \\
Fossati G., Maraschi, L., Celotti, A.,
                 Comastri, A. and Ghisellini, G., 1998, MNRAS, 299, 433\\
Le Bohec, S., et al. 1998, Nucl. Instr. Meth. A416, 425\\
Protheroe, R.J., et al., 1997, Proc. 25$^\mathrm{th}$ ICRC 
               (Durban), vol. 8, p. 317\\
Punch, M. 1997, Proc. 25$^\mathrm{th}$ ICRC 
               (Durban), vol. 3, p. 253\\
Quinn, J., et al., 1996, ApJ, 456, L83\\
Renault C., Renaud N. and Henri G. 1998,  the 19th Texas Symposium on
                     Relativistic Astrophysics and Cosmology, held in
                     Paris, France, Dec. 14-18, 1998. Eds.: J. Paul, T.
                     Montmerle, and E. Aubourg (CEA Saclay)\\
Samuelson, F., et al., 1998, ApJ, 501, 17
\end{document}